\documentclass[%
 aip,
 pop,%
 amsmath,amssymb,
 reprint,%
]{revtex4-1}

\usepackage{graphicx}
\usepackage{dcolumn}
\usepackage{bm}

\begin{document}

\preprint{AIP/123-QED}

\title[The role of guide field in magnetic reconnection driven by island coalescence]{The role of guide field in magnetic reconnection driven by island coalescence}

\author{A. Stanier}
\email{stanier@lanl.gov}
\author{W. Daughton}
\author{Andrei N. Simakov}
\author{L. Chac\'on}
\author{A. Le}
\affiliation{Los Alamos National Laboratory, Los Alamos, New Mexico 87545, USA}
\author{H. Karimabadi}
\affiliation{12837 Caminito del Canto, Del Mar, California 92014, USA}
\author{Jonathan Ng}
\author{A. Bhattacharjee}
\affiliation{Center for Heliophysics, Princeton Plasma Physics Laboratory, Princeton, New Jersey 08540, USA}

\date{\today}

\begin{abstract}

A number of studies have considered how the rate of magnetic reconnection scales in large and weakly collisional systems by the modelling of long reconnecting current sheets. However, this set-up neglects both the formation of the current sheet and the coupling between the diffusion region and a larger system that supplies the magnetic flux. Recent studies of magnetic island merging, which naturally include these features, have found that ion kinetic physics is crucial to describe the reconnection rate and global evolution of such systems. In this paper, the effect of a guide field on reconnection during island merging is considered. In contrast to the earlier current sheet studies, we identify a limited range of guide fields for which the reconnection rate, outflow velocity, and pile-up magnetic field increase in magnitude as the guide field increases.  The Hall-MHD fluid model is found to reproduce kinetic reconnection rates only for a sufficiently strong guide field, for which ion inertia breaks the frozen-in condition and the outflow becomes Alfv\'enic in the kinetic system. The merging of large islands occurs on a longer timescale in the zero guide field limit, which may in part be due to a mirror-like instability that occurs upstream of the reconnection region.

\end{abstract}
\maketitle

\section{Introduction}

Magnetic reconnection is the process of changing magnetic field-line connectivity in highly conducting plasmas,~\cite{priest00,zweibel09} and is usually associated with the conversion of a large amount of magnetic energy into plasma kinetic energy. The question of how fast magnetic flux can be reconnected in large and weakly collisional systems is important in space weather modelling,~\cite{angelopoulos08,schindler06} magnetic confinement fusion devices,~\cite{wesson90,yamada94,chapman10,stanier13} and the understanding of many astrophysical phenomena. A significant number of non-linear simulation studies have considered this question with both fully kinetic and a number of reduced fluid models. One key result from these studies was that the Hall-MHD fluid model, in which ion inertia is responsible for setting the ion diffusion region thickness, was able to reproduce the reconnection rates of fully kinetic simulations.~\cite{birn01} The reconnection rate did not depend on the detailed kinetic physics of electrons or ions,~\cite{birn01,hesse99,shay01,horiuchi94} since this physics is absent in the Hall-MHD model.

The majority of these studies used the Harris-sheet set-up,~\cite{harris62} in which a pre-formed current sheet is often highly unstable due to its kinetic scale thickness and macro-scale length. Further, such a system does not include important features that are present in real reconnecting systems, such as the formation of the current sheet, and its coupling to a large scale system that supplies the magnetic flux to the reconnection site. The magnetic island coalescence problem~\cite{finn77,biskamp80,dorelli03,knollchacon06a,pritchett07,wan08,karimabadi11} naturally includes such features, since the flux is supplied by two macro-scale magnetic islands and a current sheet forms self-consistently between the islands as they collide. Such magnetic islands are two-dimensional representations of magnetic flux ropes, a fundamental building block of magnetised plasmas.~\cite{sun10,daughton11,gekelman16} The interaction of islands, or flux-ropes, is thought to be a common process in the solar corona, the solar wind and the Earth's magnetosphere. In particular, island coalescence has been inferred both by indirect observations of coronal mass ejections,~\cite{gopalswamy02,song12} and by in-situ detection in the magnetotail.~\cite{wang16}

Recent simulation studies~\cite{karimabadi11,stanier15prl,ng15} have revisited the conclusions from earlier current layer studies using the island coalescence problem set-up. It was found that the Hall-MHD model was unable to reproduce the geometry of the ion diffusion region, the upstream magnetic field `pile-up' strength, the outflow velocity, or the reconnection rate including its dependence on ion temperature or system size.~\cite{stanier15prl} Further, these differences lead to different global evolution of the system - for Hall-MHD the islands merge when they first approach, but for the fully kinetic model the islands bounce off each other to give a much longer timescale for merging. Ion kinetic physics, and in particular the anisotropic and agyrotropic nature of the ion pressure tensor, was identified as the crucial physics missing from the Hall-MHD model. A hybrid model with mass-less fluid electrons and kinetic ions was able to reproduce many of the full kinetic results.

In the present study, the sensitivity of these results to the addition of a finite guide field is considered. Such a guide field is not reconnected, but it can modify the orbits of ions and electrons around the reconnection site. It is found that the reconnection rate, pile-up magnetic field, and outflow velocity increase with guide field for a limited range of guide field strengths that are applicable to the Earth's magnetosphere. The Hall-MHD model is able to reproduce the fully kinetic reconnection rate only in the limit of strong guide fields, when the ion pressure tensor physics becomes sufficiently localised around the x-point and the ion inertia is responsible for setting the ion diffusion region thickness. For large systems with kinetic ions, it is found that the time for the islands to fully coalesce is significantly shorter when there is a guide field present, compared with the zero guide field case.

The paper is organised as follows. In Sec.~\ref{models}, a description of the different plasma models used in this study is presented, and Sec.~\ref{probsetup} gives the detailed parameters and initial conditions for the island coalescence simulations. The detailed role of the guide field in small systems is discussed in Sec.~\ref{smallguide}, and then additional effects in large systems are given in Sec.~\ref{largeguide}. Finally, the important results are summarised and conclusions are drawn in Sec.~\ref{conclusions}.

\section{\label{models}Description of plasma models employed}

A key aim of this study is to evaluate the fidelity of several commonly used reduced physics models for the island coalescence problem with a guide field, by comparing them against fully kinetic Particle-In-Cell (PIC) simulations. A description of these models follows, along with a very brief description of the main numerical methods used.

\subsection{Fully kinetic particle-in-cell}

The fully kinetic simulations are carried out using the high performance electromagnetic relativistic particle-in-cell (PIC) code VPIC.~\cite{bowers08,bowers09} The electric and magnetic fields are advanced with the Maxwell-Ampere and Maxwell-Faraday equations
\begin{equation}\label{ampere}\frac{1}{c^2}\frac{\partial \boldsymbol{E}}{\partial t} = \boldsymbol{\nabla}\times \boldsymbol{B} - \mu_0 \boldsymbol{j},\end{equation}
\begin{equation}\label{faraday}\frac{\partial \boldsymbol{B}}{\partial t} = -\boldsymbol{\nabla}\times \boldsymbol{E}.\end{equation}
Here $\boldsymbol{E}$ is the electric field, $\boldsymbol{B}$ is the magnetic field, $\boldsymbol{j}$ is the current density, $c$ is the speed of light and $\mu_0$ is the vacuum permittivity.
The particles are advanced using
\begin{equation}\label{partx}\frac{d \boldsymbol{x}_{s,m}}{d t} = c \boldsymbol{u}_{s,m}/\gamma_{s,m},\end{equation}
\begin{equation}\label{partu}\frac{d \boldsymbol{u}_{s,m}}{d t} = \frac{q_s}{m_s c} \left[\boldsymbol{E} + c\boldsymbol{u}_{s,m}\times \boldsymbol{B}/\gamma_{s,m}\right],\end{equation}
where $\boldsymbol{u}_{s,m} = \gamma_{s,m}\boldsymbol{v}_{s,m}/c$ is the normalised momentum of the particle $m$ from species $s$, $\boldsymbol{x}_{s,m}$ is the particle position, $q_s$ and $m_s$ are the charge and rest-mass for particles of species $s=i,e$ for ions and electrons, and $\gamma_{s,m}=\sqrt{1+u_{s,m}^2}$.

VPIC advances fields on a Yee-mesh with a finite-difference time-domain method and particles are pushed with a Boris algorithm modified to use a sixth order rotation angle approximation. The electromagnetic fields are interpolated to the particles using an energy conserving scheme, and the current density is gathered using the charge conserving Villasenor-Buneman method. See Refs.~[\onlinecite{bowers08}],~[\onlinecite{bowers09}] and references therein for further details.

\subsection{Hybrid model}

The hybrid and Hall-MHD models solve for the magnetic field advance using Eq.~(\ref{faraday}), but calculate the current density as $\boldsymbol{j} = \boldsymbol{\nabla}\times \boldsymbol{B}/\mu_0$. These models also enforce charge neutrality as given for a hydrogen plasma by equal ion and electron number densities, $n = n_i = n_e$, and assume mass-less electrons, $m_e = 0$. The electric field is then found from Ohm's law

\begin{equation}\label{ohms} \boldsymbol{E} = -\boldsymbol{v}\times \boldsymbol{B} + \frac{1}{ne}\left(\boldsymbol{j}\times \boldsymbol{B} - \boldsymbol{\nabla}p_e\right) + \eta \boldsymbol{j} - \eta_H \nabla^2 \boldsymbol{j},\end{equation}
where $p_e$ is the scalar electron pressure, $\eta$ is the resistivity, $\eta_H$ is the hyper-resistivity and $e=|q_e|$ is the elementary charge.

The hybrid model treats ions kinetically using the PIC method, while electrons are treated as a mass-less fluid. The ions are advanced with Eqs.~(\ref{partx}) and~(\ref{partu}) in the non-relativistic limit ($\gamma_{s,m} \rightarrow 1$), and the density $n$ and bulk ion velocity $\boldsymbol{v}$ are calculated from velocity moments of the particle distribution. The electrons are assumed isothermal $p_e = T_{e0} n$, where $T_{e0}$ is the initial temperature. Eq.~(\ref{faraday}) is solved with the Runge-Kutta method and particles are pushed using the standard Boris method. See Ref.~[\onlinecite{karimabadi11code}] and references therein for further details.

\subsection{Hall-MHD model}

The Hall-MHD fluid model solves additional fluid conservation equations for plasma number density, total momentum, and the isotropic total pressure
\begin{equation}\label{continuity}\frac{\partial n}{\partial t} + \boldsymbol{\nabla} \cdot \left(n\boldsymbol{v}\right) = 0,\end{equation}
\begin{align}\label{momentum}\frac{ \partial \left(m_in\boldsymbol{v}\right)}{\partial t} &+ \boldsymbol{\nabla} \cdot [m_i n\boldsymbol{v}\boldsymbol{v} - \boldsymbol{B}\boldsymbol{B}/\mu_0 + \nonumber \\
&\tensor{\boldsymbol{I}} \left(p + B^2/2\mu_0\right) + \tensor{\boldsymbol{\Pi}}] = 0,\end{align}
\begin{equation}\label{pressure}\frac{\partial p}{\partial t} + \boldsymbol{\nabla} \cdot \left(\boldsymbol{v}^* p\right) + \left(\gamma -1\right) p \boldsymbol{\nabla} \cdot \boldsymbol{v}^* = \left(\gamma -1\right)\left(Q - \boldsymbol{\nabla} \cdot \boldsymbol{q}\right) = 0.\end{equation}
Here, a constant temperature ratio approximation $\alpha = T_i/T_e=\textit{const.}$ is used to give a single equation for the total pressure $p=p_i + p_e = n\left(1+\alpha\right)T_e$ carried by the effective velocity $\boldsymbol{v}^* = \boldsymbol{v} - \boldsymbol{j}/[ne(1+\alpha)]$. Simple heat conduction and heating terms are used, $\boldsymbol{q} = -\kappa \boldsymbol{\nabla} T_e$ and $Q = \eta j^2 + \eta_H \left(\boldsymbol{\nabla}\boldsymbol{j}\right)^2 -\tensor{\boldsymbol{\Pi}} : \boldsymbol{\nabla}\boldsymbol{v}$, respectively, where the ion viscous tensor is given by $\tensor{\boldsymbol{\Pi}} = - \mu \boldsymbol{\nabla}\boldsymbol{v}$. $\mu$ is the viscosity, $\kappa$ is the heat conductivity, and $\gamma$ is the heat capacity ratio.

The system of equations~(\ref{faraday}),~(\ref{ohms}),~(\ref{continuity})-(\ref{pressure}) is advanced using a fully implicit, preconditioned Newton-Krylov method.~\cite{chacon08a,chacon08b} Spatial discretisation is done using a cell centred finite volume scheme,~\cite{chacon04} and the second order BDF-2 method is used for time advance.

\subsection{10-moment model}

The 10-moment model~\cite{wang15,ng15} solves the equations

\begin{equation}\frac{\partial n_s}{\partial t} + \boldsymbol{\nabla} \cdot \left(n_s \boldsymbol{v}_s\right) = 0,\end{equation}

\begin{equation}m_s\left[\frac{\partial n_s \boldsymbol{v}_s}{\partial t} + \boldsymbol{\nabla} \cdot \left(n_s \boldsymbol{v}_s\boldsymbol{v}_s\right)\right] + \boldsymbol{\nabla}\cdot \tensor{\boldsymbol{P}}_s = q_s n_s \left(\boldsymbol{E} + \boldsymbol{v}_s \times \boldsymbol{B}\right),\end{equation}

\begin{align}\frac{\partial \tensor{\boldsymbol{P}}_s}{\partial t} &+& \boldsymbol{\nabla} \cdot \left(\boldsymbol{v}_s \tensor{\boldsymbol{P}}_s\right) + \left(\tensor{\boldsymbol{P}}_s\cdot \boldsymbol{\nabla} \boldsymbol{v}_s\right) + \left(\tensor{\boldsymbol{P}}_s\cdot \boldsymbol{\nabla} \boldsymbol{v}_s\right)^T \nonumber \\
&+& \boldsymbol{\nabla} \cdot \tensor{\boldsymbol{Q}}_s = \frac{q_s}{m_s} \left(\tensor{\boldsymbol{P}}_s\times \boldsymbol{B} - \boldsymbol{B} \times \tensor{\boldsymbol{P}}_s\right),\end{align}
for $s=i,e$, in addition to Eqs.~(\ref{ampere}) and~(\ref{faraday}) using the current calculated with $\boldsymbol{j} = \sum_s n_s q_s \boldsymbol{v}_s$. Here, $\tensor{\boldsymbol{P}}_s$ and $\tensor{\boldsymbol{Q}}_s$ are the pressure and heat flux tensors for the species $s$ in its rest-frame. 

Currently, the set of moment equations is truncated by the simple closure~\cite{hesse95,wang15,ng15}
\begin{equation}\boldsymbol{\nabla} \cdot \tensor{\boldsymbol{Q}}_s = v_{Ts} k_s \left(\tensor{\boldsymbol{P}}_s - p_s\tensor{\boldsymbol{I}}\right),\end{equation}
where $p_s=\textrm{trace}[\boldsymbol{P}_s]/3$ is the isotropic pressure, and $v_{Ts}$ is the thermal velocity. This relaxes the pressure tensor to an isotropic pressure at a rate given by $|k_s|v_{Ts}$, allowing deviations from isotropy in the ion and electron pressure tensors at length scales less than $1/|k_i|$ and $1/|k_e|$ respectively. These $k_s$ are currently treated as free parameters, but work is being done to close the set of equations in the manner of Ref.~[\onlinecite{hammettperkins90}].

These equations are solved using a second order locally implicit operator splitting approach, where the hyperbolic part uses a dimensionally split finite-volume wave propagation scheme. More details can be found in Refs.~[\onlinecite{ng15}] and~[\onlinecite{wang15}].

\section{\label{probsetup}Problem set-up}

The simulations described in this paper are initialised with the exact Vlasov magnetic island equilibrium.~\cite{fadeev65,schindler06} The magnetic field is given by $\boldsymbol{B} = \boldsymbol{\nabla}\times \left(A_z \boldsymbol{\hat{z}}\right) + B_g \boldsymbol{\hat{z}}$, where $B_g$ is the guide field strength and $A_z$ is the out-of-plane magnetic potential given by
\begin{equation}A_z=B_0\lambda \ln{\left[\cosh{\left(x/\lambda\right)} + \epsilon \cos{\left(y/\lambda\right)}\right]}.\end{equation}
Here, $\epsilon = 0.4$ and $B_0$ is the asymptotic magnetic field. 

For equilibrium, it is required that the sum of initial ion and electron temperatures is $T_{i0} + T_{e0} = B_0^2/\left(2\mu_0 n_0 k_B\right)$, where $n_0$ is the reference density equal to the density enhancement at the centre of the current layer in the limit $\epsilon = 0$, and $k_B$ is the Boltzmann constant. In this paper the ratio of initial temperatures used is $T_{i0}/T_{e0} = 1$. The density profile is given by
\begin{equation} n = n_b+ \frac{n_0\left(1-\epsilon^2\right)}{\left[\cosh{\left(x/\lambda\right)} + \epsilon \cos{\left(y/\lambda\right)}\right]^2},\end{equation}
where $n_b = 0.2 n_0$ is a background density. The size of the simulation domain is $x\in \left[-\pi \lambda,\pi \lambda\right], \, y\in  \left[-2\pi \lambda,2\pi \lambda\right]$. For all codes we use perfectly conducting boundaries that are reflecting for particles at $x=\pm \pi \lambda$, and periodic boundaries in the $y$-direction. We use a sinusoidal perturbation of magnitude $0.1B_0$ to start the merging.

Additional parameters specific to each model are as follows. For fully kinetic PIC simulations we use a ratio of electron plasma to gyro-frequency $\omega_{pe}/\Omega_{ce} = 2$, and mass-ratio $m_i/m_e = 25$. For the hybrid model, we use a ratio of the ion frequencies $\omega_{pi}/\Omega_{ci} = 2000$, a resistivity $\eta=10^{-5}\mu_0 d_i v_{A0}$, and a hyper-resistivity $\eta_H = 10^{-4}\mu_0 d_i^3 v_{A0}$, where $d_i = \sqrt{m_i/(n_0e^2\mu_0)}$ is the ion skin depth and $v_{A0} = B_0/\sqrt{m_in_0\mu_0}$ is the Alfv\'en speed. This value of hyper-resistivity is an order of magnitude smaller than in Ref.~[\onlinecite{stanier15prl}] for the hybrid model. The value $\eta_H=10^{-3}\mu_0 d_i^3 v_{A0}$ was found to be sufficient in the zero guide field case, but with finite guide field the ion diffusion region is thinner (see below) and a smaller $\eta_H$ is necessary to ensure good separation between ion and electron scales. This change in $\eta_H$ gives only a $2\%$ reduction of the peak reconnection rate for the zero guide field run with $\lambda=5d_i$ reported in Ref.~[\onlinecite{stanier15prl}].

For Hall-MHD, we use a fixed temperature ratio $\alpha =1$, ion viscosity $\mu = 10^{-2}m_in_0d_iv_{A0}$, resistivity $\eta=10^{-5}\mu_0 d_i v_{A0}$, hyper-resistivity $\eta_H = 10^{-4}\mu_0 d_i^3 v_{A0}$, and heat conduction $\kappa = 10^{-4}n_0 d_i v_{A0}$. Finally, for 10-moment simulations we use $\omega_{pe}/\Omega_{ce} = 2$, $m_i/m_e = 25$, $k_e = 1/d_e$ and $k_i = 1/3d_i$, where $d_e = \sqrt{m_e/m_i}d_i$ is the electron-skin depth. As for Harris-sheet geometry, we find that the precise value of the dissipation and $m_i/m_e$ do not influence the reconnection rate, provided there is a sufficient separation between ion and electron scales.

To compare with previous studies,~\cite{karimabadi11,stanier15prl,ng15} we normalise the magnetic fields by the maximum value of the initial in-plane field in a line joining the island O-points, $B_m = 0.353B_0$. Velocities are normalised by $v_{Am} = B_m v_{A0}/B_0$, and lengths are normalised by $d_i$. Times are given in terms of the global Alfv\'en crossing time $\tau_A = 4\pi \lambda/v_{A0}$ to easily compare between different system sizes. Finally, reconnection rates are measured as $E_R =  \partial_t \left[A_{zX} - A_{zO}\right]/(v_{Am}B_m)$, where $A_{zX/O}$ is the $A_z$ evaluated at the $X/O$ magnetic null point, and the average reconnection rate $<E_R>$ is the value of $E_R$ time averaged over $1.5\tau_A$.

In this study, we vary the strength of the guide field $B_g$, and the half-thickness of the equilibrium current layer $\lambda$. While varying $\lambda$, we keep $\epsilon$ fixed and the domain size proportional to $\lambda$. Thus, variations in $\lambda$ are equivalent to varying the system size with respect to $d_i$.

\section{\label{smallguide}Effect of guide field on reconnection rate}

\subsection{Reconnection rate}

\begin{figure}
\includegraphics[scale=0.8]{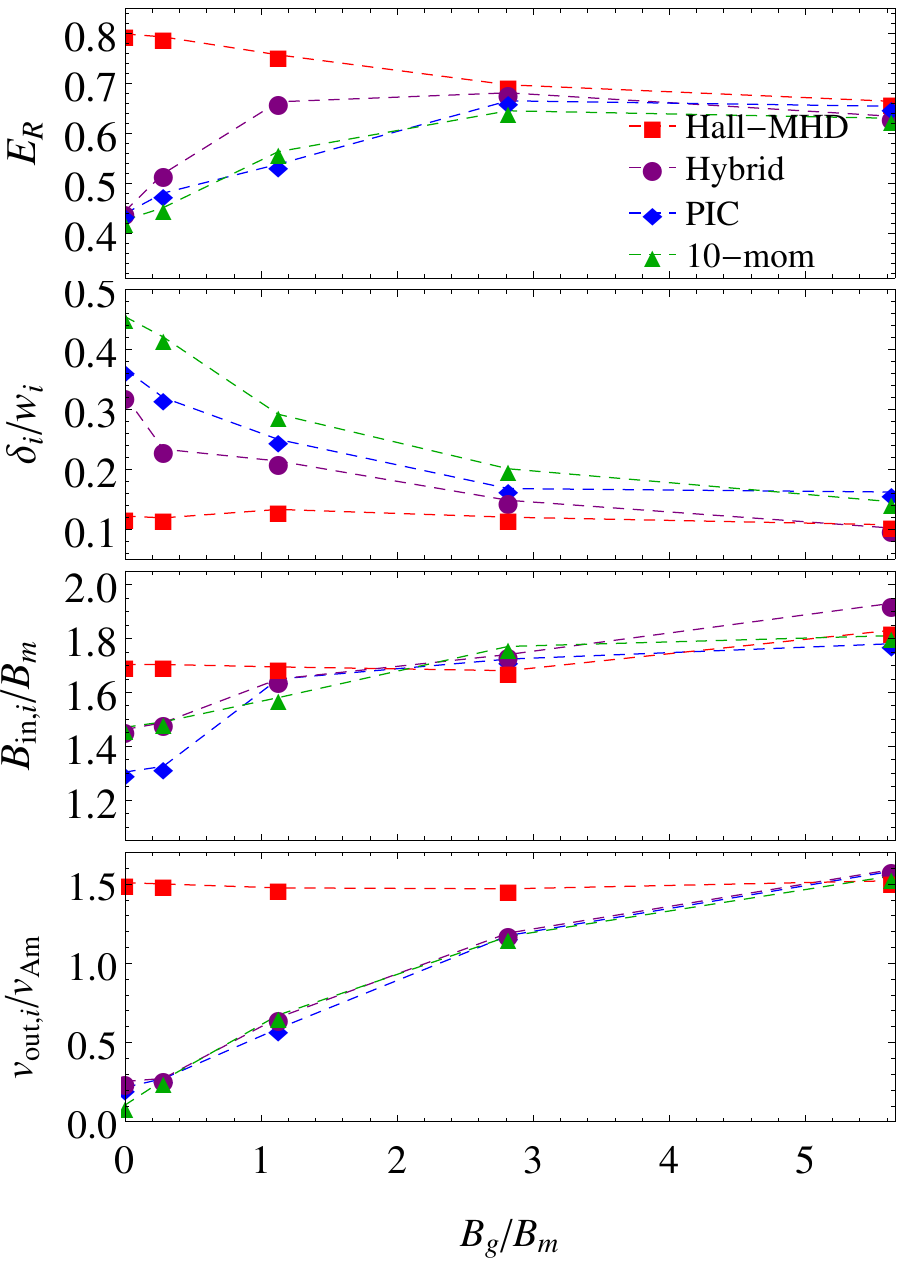}
\caption{\label{fig:quantities-guide} Dependence of the peak reconnection rate $E_R$, aspect ratio of the ion diffusion region $\delta_i/w_i$, normalised pile-up magnetic field strength $B_{\textrm{in},i}/B_m$, and normalised outflow velocity $v_{\textrm{out},i}/v_{Am}$ on the normalised guide magnetic field strength $B_g/B_m$.}
\end{figure}

Figure~\ref{fig:quantities-guide} shows how the peak reconnection rate $E_R$, the aspect ratio of the ion diffusion region $\delta_i/w_i$, the normalised pile-up magnetic field $B_{\textrm{in},i}/B_m$, and the normalised peak outflow velocity $v_{\textrm{out},i}/v_{Am}$ scale with normalised guide field $B_g/B_m$ for runs with $\lambda = 5d_i$ for the different plasma models. Here, $\delta_i$ and $w_i$ are measured as the full-width half-maxima of the non-ideal electric field $E'_z = \boldsymbol{\hat{z}}\cdot \left(\boldsymbol{E} + \boldsymbol{v}\times \boldsymbol{B}\right)$ in cuts across the ion diffusion region with $x=0$ and $y=0$ respectively. $B_{\textrm{in},i}$ is the maximum value of the upstream magnetic field, and $v_{\textrm{out},i}$ is the maximum outflow velocity. For the Harris-sheet problem, $B_g$ is often normalised by the asymptotic field $B_0$. For our normalisaton, the value of $B_g = 2.83B_m = 1B_0$.

As discussed in Ref.~[\onlinecite{stanier15prl}], for $B_g = 0$ the Hall-MHD model is unable to reproduce the equivalent fully kinetic PIC values of the quantities plotted in Fig.~\ref{fig:quantities-guide}. The simulations which retain the full physics of kinetic ions (hybrid and PIC) have a broader ion diffusion regions and show significantly reduced rates, pile-up and outflow velocities with respect to Hall-MHD. In particular, the PIC simulation with $B_g = 0$ has a  peak outflow velocity 7 times smaller than Hall-MHD, an aspect-ratio $3.2$ times larger, and the pile-up at $80\%$ of the Hall-MHD value. To illustrate how these differences modify the rate $E_R$, it is useful to consider a quasi-steady and rectangular ion diffusion region (iDR) with constant inflow and outflow velocities along the edges. The reconnection rate is then set by the inflowing flux to the iDR $v_{\textrm{in},i}B_{in,i}$, where the inflow velocity is constrained by the continuity equation $v_{\textrm{in},i} \approx v_{\textrm{out},i} (\delta_i/w_i) (n_{\textrm{out}}/n_{\textrm{in}})$. For the numbers above, and with little change in $n_{\textrm{out}}/n_{\textrm{in}}$ between Hall-MHD and PIC, this simple estimate gives a factor of $2.7$ reduction in the rate for PIC that is broadly consistent with the $\approx 2$ times smaller $E_R$ in Fig.~\ref{fig:quantities-guide}.

For the range of $B_g$ in Fig.~\ref{fig:quantities-guide}, the plotted quantities for the Hall-MHD simulations have weak guide field dependence. In particular, the peak rate $E_R$ decreases by only $13\%$ from $B_g=0$ to $B_g/B_m = 2.83$. In contrast, the peak rates increase by $51\%$ ($53\%$) for PIC (hybrid) across this range to give $E_R = 0.665$ ($0.681$) for $B_g/B_m = 2.83$ that are comparable to Hall-MHD ($E_R = 0.697$). As far as we are aware, such an increase in the reconnection rate with guide field has been noted previously only in asymmetric reconnection,~\cite{hesse13asym,pritchett08asym} where there are different thermal and magnetic pressures on either side of the current sheet. Symmetric current layer studies typically find that the reconnection rate is independent of $B_g$ for $B_g \leq B_0$, then begins to decrease~\cite{pritchett043D, ricci04} as $B_g$ increases until it flattens again once the sound Larmor radius $\rho_s = c_s/\Omega_{ci}$, defined with the sound speed $c_s$ and ion cyclotron frequency $\Omega_{ci}$, falls below the electron-skin depth.~\cite{liu14,stanier15b} For larger guide fields, such as $B_g/B_m=5.66$ shown in Fig.~\ref{fig:quantities-guide}, $E_R$ begins to gradually decrease in the same way for Hall-MHD, hybrid and PIC. However, the focus of the present study is on the range of weak guide fields for which there is qualitatively different behaviour from the extended current layer simulations ($B_g/B_m \leq 2.83$). This regime has application in magnetospheric plasmas.

In addition to the increase in $E_R$, hybrid and fully kinetic PIC simulations show strong dependences on $B_g$ for $\delta_i/w_i$, $B_{\textrm{in},i}/B_m$ and $v_{\textrm{out},i}/v_{Am}$, but all the quantities are close to the Hall-MHD results at $B_g/B_m \ge 2.83$.

Finally, as seen in Fig.~\ref{fig:quantities-guide}, the 10-moment model is able to reproduce the same trends as fully kinetic PIC in all of the plotted quantities for the range of guide fields considered. In particular, both the rates $E_R$ and outflow velocities $v_{\textrm{out,i}}/v_{Am}$ are in excellent agreement between the 10-moment and fully kinetic PIC simulations.

\subsection{Ion diffusion region physics}

\begin{figure}
\includegraphics[scale=0.8]{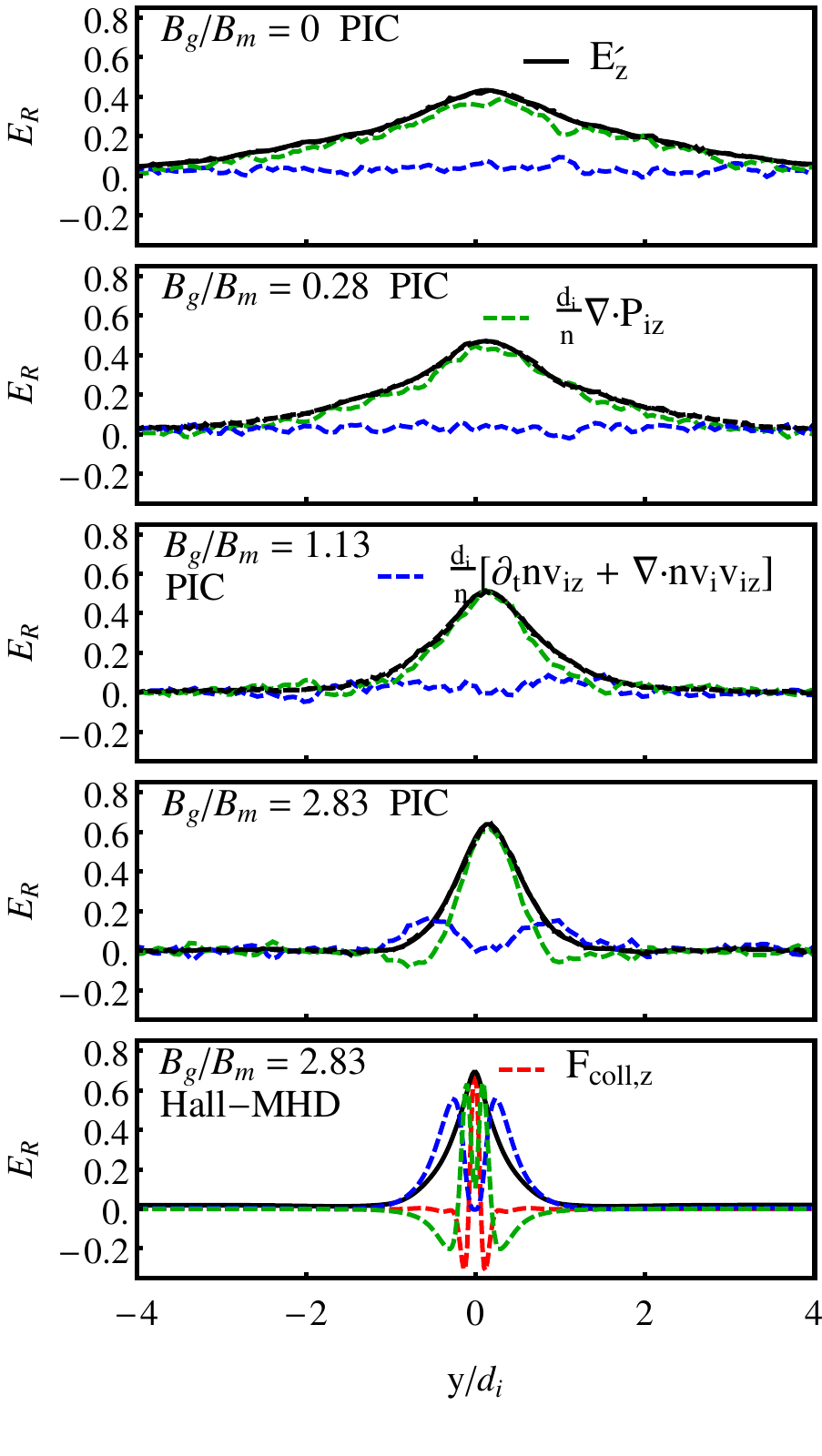}
\caption{\label{fig:idr-thickness-guide} Terms from Eq.~(\ref{ionohms}) in cuts across the ion diffusion region from VPIC simulations with $\lambda=5d_i$ and $B_g/B_m = 0, 0.28, 1.13$ and $2.83$, respectively, and from a Hall-MHD simulation with $\lambda=5d_i$ and $B_g/B_m = 2.83$.}
\end{figure}

The decrease in the aspect ratio $\delta_i/w_i$ with the increase in $B_g$, for all of the models except Hall-MHD, follows from a significant decrease in the thickness of the ion diffusion region $\delta_i$. The length of this region $w_i$ also decreases with guide-field but by a smaller amount (not shown). Figure~\ref{fig:idr-thickness-guide} shows the contributions to the non-ideal electric field $E_z'$ in 1D cuts across the ion diffusion region in the inflow direction ($x=0$) for fully kinetic PIC simulations (top four panels) with $\lambda = 5 d_i$ and varying guide field. The contributions to $E_z'$ are given by the ion momentum equation, which in a normalised form is given by
\begin{equation}\label{ionohms}E'_z = \frac{d_i}{n}\left[\partial_t\left(nv_{iz}\right) + \boldsymbol{\nabla} \cdot \left(n\boldsymbol{v}_i v_{iz}\right)\right]+ \frac{d_i}{n}\boldsymbol{\nabla}\cdot\boldsymbol{P}_{iz} + F_\textrm{coll,z},\end{equation}
where the collisionless PIC simulations only have contributions from the ion inertia (blue) and ion pressure tensor (green) terms.

As mentioned above, $\delta_i$ is measured as the full-width at half-maximum (FWHM) of the $E_z'$ curve (black) in each case. This thickness decreases from $\delta_i = 2.8 d_i$ for $B_g = 0$ to $\delta_i = 0.85d_i$ for $B_g/B_m = 2.83$. For $B_g = 0$ the ions exhibit characteristic meandering-type orbits as they cross the weak-field region,~\cite{speiser65,horiuchi94,hesse99,stanier15prl} giving rise to strong gradients in the off-diagonal elements of the ion pressure tensor. The extent of the ion orbits is reduced by the addition of a finite guide field, and the orbits may become chaotic if the gyro-radius becomes comparable to the radius of magnetic curvature, $R_c$.~\cite{buchner89} For a strong enough guide field, the gyro-radius will fall below both the meandering length and $R_c$, and the ions start to become magnetised. 

Fig.~\ref{fig:idr-thickness-guide} shows a transition at $B_g = 2.83B_m$ to ion inertia (blue) supporting $E_z'$ at the edges of the iDR, and thus setting the thickness. The pressure tensor term (green) continues to balance $E_z'$ close to the stagnation point ($y=0$) where ion inertia is small provided the diffusion region is quasi-steady, $\partial_t (nv_{iz}) \approx 0$. For the island coalescence problem this quasi-steady phase is short in duration and occurs at the time of the peak reconnection rate.~\cite{simakov06} 

For comparison, the contributions to $E_z'$ are also plotted from the Hall-MHD simulation with $\lambda = 5d_i$ and $B_g = 2.83B_m$ in the bottom panel of Fig.~\ref{fig:idr-thickness-guide}. Here, as in Ref.~[\onlinecite{stanier15prl}], $\boldsymbol{P}_{iz} = -\mu \boldsymbol{\nabla}v_{iz}$ is a collisional viscosity (green curve) and $F_{\textrm{coll},z} = \eta j_z -\eta_H \nabla^2 j_z$ is the collisional friction (red curve), which is almost entirely from the hyper-resistive term. The thickness of the ion-diffusion region for the Hall-MHD run with $B_g/B_m = 2.83$ is $\delta_i = 0.56 d_i$ ($35\%$ smaller than the PIC value), which is reduced only slightly from the zero-guide field case ($\delta_i = 0.62 d_i$, not shown). In the Hall-MHD simulations, ion inertia supports $E_z'$ at the edges of the ion diffusion region and sets the thickness $\delta_i$ for the full range of guide-fields of Fig.~\ref{fig:quantities-guide}. In this sense, there is qualitative agreement in the physics breaking the frozen-in condition for ions between the Hall-MHD and PIC models only for the strongest guide field case, $B_g/B_m = 2.83$. For Hall-MHD, the viscous and hyper-resistive sub-layers in $E_z'$ are significantly thinner than the pressure tensor sub-layer in the $B_g=2.83B_m$ PIC result. Their thickness depends on the values of the dissipation coefficients $\mu$ and $\eta_H$; however, the reconnection rate is insensitive to their precise thickness provided they are thin in comparison to $\delta_i$.

In the absence of a guide field and for a small system size ($\lambda=5d_i$), it was demonstrated in Ref.~[\onlinecite{ng15}] that the 10-moment model was able to reproduce either the Hall-MHD or kinetic values for $E_R$ and $\delta_i$, depending on the value chosen for the parameter $k_i$. For $k_i = 1/d_e$ the pressure tensor is allowed to depart from isotropy only on scales smaller than the electron skin-depth. In this case, the ion frozen-in condition is violated at $d_i$-scale due to the ion inertial term. However, setting $k_i = 1/\delta_i$, where $\delta_i=3d_i$ is the approximate thickness of the ion diffusion region for the zero guide field PIC simulations, allows the ion pressure tensor to break the frozen-in condition on this larger scale. For $B_g = 0$ the thickness of the iDR for the 10-moment model is indeed at this larger scale, $\delta_i = 2.6d_i$, but for the stronger guide field case $B_g/B_m = 2.83$ the thickness reduces to $\delta_i = 0.952d_i$ and the contributions to $E'_z$ are in qualitative agreement with the equivalent PIC result (fourth panel of Fig.~\ref{fig:idr-thickness-guide}). This suggests that the $k_i$ parameter only determines the maximum allowable thickness of the ion diffusion region for the 10-moment model, $\delta_i \leq 1/k_i$. 

\begin{figure}
\includegraphics[scale=0.32]{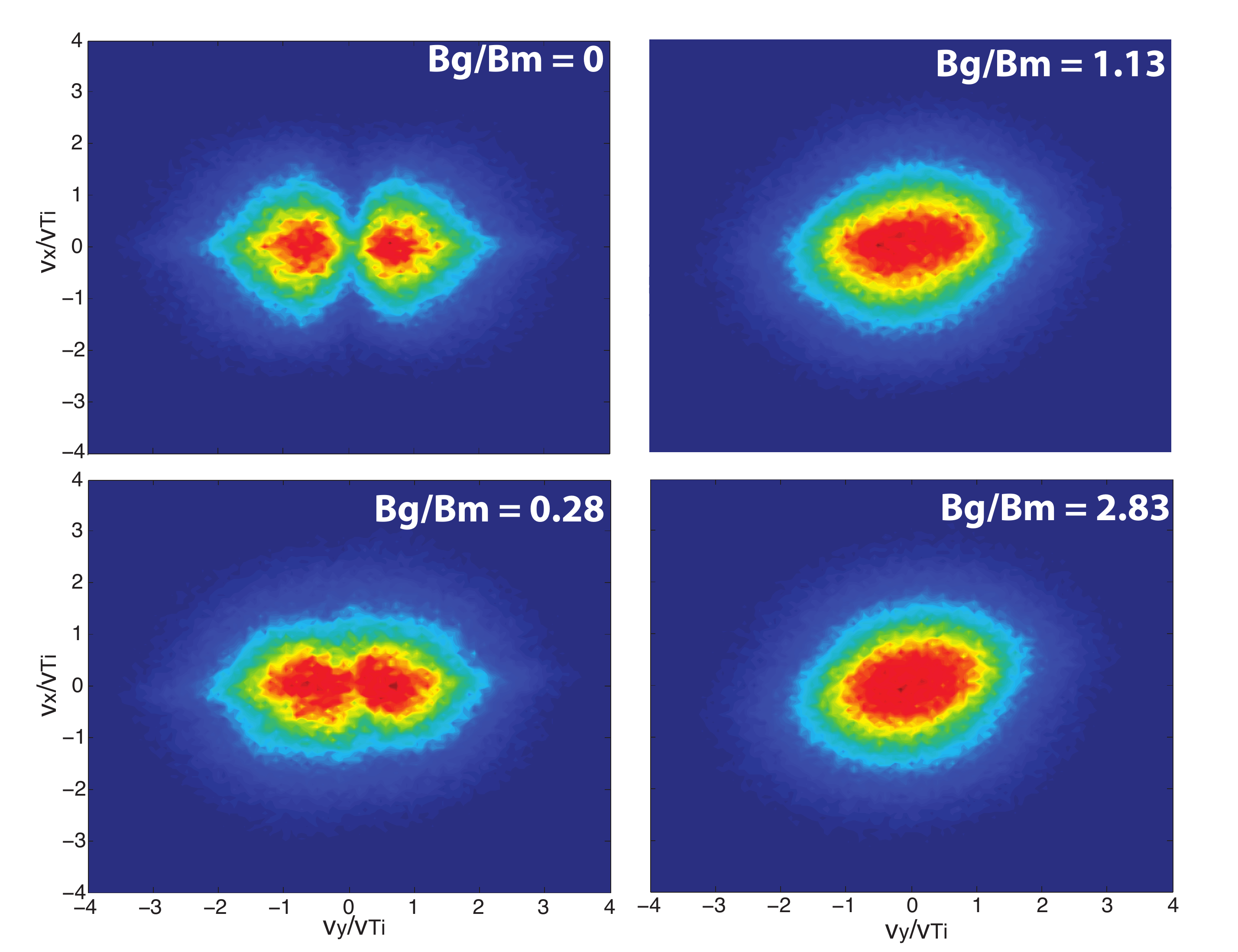}
\caption{\label{fig:distributions-guide}Distribution functions in $v_y$-$v_x$ space collected at the x-point for PIC simulations with $\lambda=5d_i$ and different normalised guide fields $B_g/B_m$.}
\end{figure}

The modification of the ion kinetic physics at the X-point with guide field can be seen in the ion distribution functions, shown in Figure~\ref{fig:distributions-guide} from the same fully kinetic PIC simulations. These distribution functions were calculated by collecting ions in a square box with $2d_e$ edge length, centred on the X-point and time averaged over $1\Omega_{ci}^{-1}$. For $B_g = 0$ there is a clear double peaked structure along the $v_y$ axis (the inflow direction), which is characteristic of the ion meandering orbits. This structure persists for weak guide-field $B_g = 0.28 B_m$, but for $B_g = 1.13 B_m$ and $B_g = 2.83 B_m$ the distribution starts to become more gyrotropic about the guide field direction. 

\subsection{Pile-up magnetic field strength and outflow velocity}

It is important to note that the reduction in $\delta_i$ with guide field alone can not explain the increase in the reconnection rate with $B_g/B_m$ from the PIC simulations in Fig.~\ref{fig:quantities-guide}. For the estimate of the rate based upon a quasi-steady rectangular ion diffusion region above, the rate would be expected to decrease as $E_R\propto \delta_i$. However, it is clear from Fig.~\ref{fig:quantities-guide} that there are changes in the other quantities that influence the reconnection rate. 

Firstly, the strength of the pile-up magnetic field $B_{in,i}/B_m$ increases with guide field for PIC, hybrid and 10-moment models until it reaches the Hall-MHD value for $B_g/B_m \approx 1.132$. The strength of this pile-up field is set by the cross-scale coupling between the global motion of the interacting islands and the micro-scale physics of the diffusion region. 

To interpret this increasing trend with guide field, it is helpful to consider previous studies of island coalescence. In resistive MHD~\cite{biskamp80} it was found that the pile-up increases as the current sheet thickness decreases (with increasing Lundquist number) until it saturates at a critical value set by pressure balance considerations.~\cite{craig93,knollchacon06a} With the inclusion of the Hall effect, the magnetic field is frozen-in to the faster electron flow below $d_i$-scale and the pile-up is reduced.~\cite{dorelli03,knollchacon06b} However, recent studies~\cite{stanier15prl,ng15} using compressible Hall-MHD, 10-moment, hybrid, and fully kinetic models have found that pile-up can still occur in larger systems just upstream of the ion diffusion region, and the strength of pile-up increases as the system-size becomes large with respect to the iDR thickness $\delta_i$. In Fig.~\ref{fig:quantities-guide} the increase in pile-up with guide field is consistent with the increase in system-size with respect to $\delta_i$, as $\delta_i$ decreases significantly over this range of guide fields for all models except Hall-MHD.

The increased pile-up gives a larger upstream Alfv\'en speed and so may increase the outflow speed, if the outflow is accelerated by the relaxing magnetic tension in newly reconnected field lines. For Hall-MHD the outflow speed is indeed comparable, see Fig.~\ref{fig:quantities-guide}, to the effective upstream Alfv\'en speed $v_{out,i} \approx v_{Am} (B_{in,i}/B_m)\sqrt{n_0/n}$ set by the pile-up field $B_{in,i}/B_m$ and upstream density $n/n_0\approx 1.4$ on the edge of the iDR. However, as mentioned above, the kinetic ion codes have an outflow velocity $\approx 7$ times smaller for $B_g=0$. Also, unexpectedly, the ion outflow speed is found to increase significantly with guide field.

\begin{figure}
\includegraphics[scale=0.7]{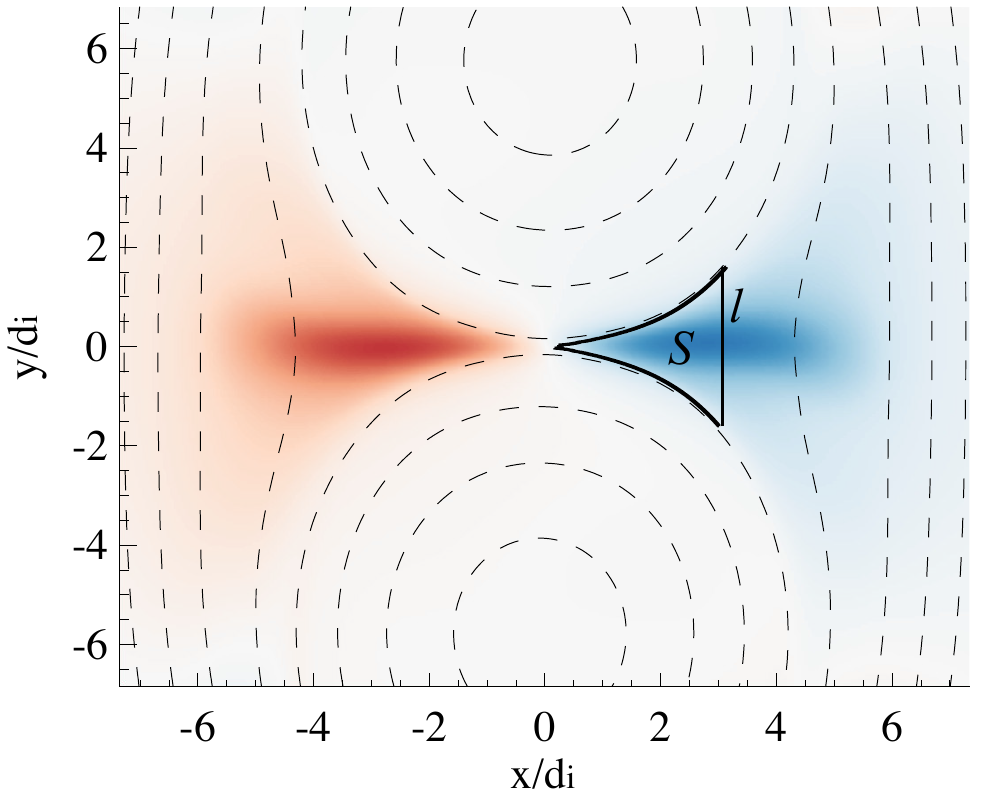}
\caption{\label{fig:wedge}Wedge shaped integration region $S$ bounded by closed contour $l$ (solid line). Dashed lines are contours of the magnetic flux and colour scale shows ion outflow velocity component $v_x$. }
\end{figure}

To understand this effect we consider the momentum balance through a wedge shaped region embedded within the outflow, following the method used in Ref.~[\onlinecite{le14}]. The wedge region shown in Fig.~\ref{fig:wedge} is found by integrating along the separatrix field-lines on one side of the X-point to $y=\pm y_{\textrm{max}}$, then the contour is closed by joining the ends with a straight line intersecting the ion outflow jet.

Provided the exhaust is quasi-steady, 
\begin{equation}\label{contourint}\int_S dS \,\boldsymbol{\nabla} \cdot \tensor{\boldsymbol{T}} \cdot \boldsymbol{\hat{x}} = \oint_l dl\, \boldsymbol{\hat{n}}\cdot \tensor{\boldsymbol{T}} \cdot \boldsymbol{\hat{x}}=0,\end{equation}
where $l$ is the closed contour bounding the wedge region of area $S$, $\boldsymbol{\hat{n}}$ is the outward normal unit vector at a point on the contour line,  
\begin{equation}\tensor{\boldsymbol{T}}=m_i n \boldsymbol{v} \boldsymbol{v} + m_e n \boldsymbol{v}_e \boldsymbol{v}_e + p_T\tensor{\boldsymbol{I}} + \tensor{\boldsymbol{\Pi}}_T - \boldsymbol{B}\boldsymbol{B}\end{equation}
is the total momentum tensor, $p_T = p_{\perp i} + p_{\perp e} + B^2/2\mu_0$ is the total of the perpendicular thermal and magnetic pressures, and $\tensor{\boldsymbol{\Pi}}_T = (p_{\parallel i} + p_{\parallel e} - p_{\perp i} - p_{\perp e})\boldsymbol{\hat{b}}\boldsymbol{\hat{b}} + \tensor{\boldsymbol{\Pi}}^{ng}_i + \tensor{\boldsymbol{\Pi}}^{ng}_e$ is a viscous-like stress tensor that contains anisotropic pressure and non-gyrotropic terms, $p_{\parallel i,e} =  \boldsymbol{\hat{b}}\cdot \tensor{\boldsymbol{P}}_{i,e} \cdot \boldsymbol{\hat{b}}$, $p_{\perp i,e} = (\textrm{trace}[\tensor{\boldsymbol{P}}_{i,e}]-p_{\parallel i,e})/2$, and $\tensor{\boldsymbol{\Pi}}^{ng}_{i,e} = \tensor{\boldsymbol{P}}_{i,e} - p_{\parallel i,e} \boldsymbol{\hat{b}} \boldsymbol{\hat{b}} - p_{\perp i,e}(\tensor{\boldsymbol{I}}-\boldsymbol{\hat{b}} \boldsymbol{\hat{b}})$. The electron inertial contribution is very small compared to the ion contribution in all results presented.

\begin{figure}
\includegraphics[scale=1.0]{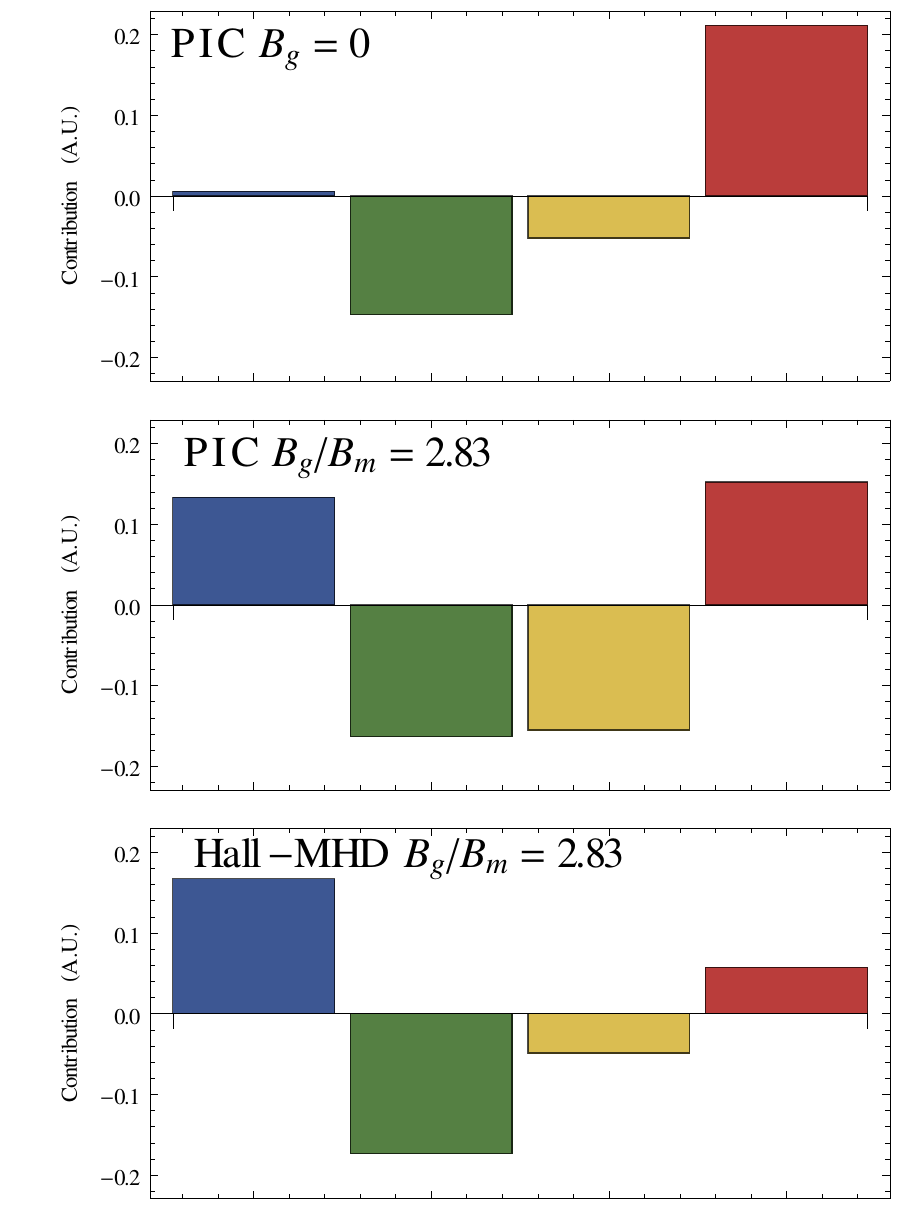}
\caption{\label{fig:wedge-contours}Contributions to Eq.~(\ref{contourint}), from $m_i n\boldsymbol{v}v_{x} + m_e n\boldsymbol{v}_ev_{ex}$ (blue), $-\boldsymbol{B}B_x$ (green), $p_T\boldsymbol{\hat{x}}$ (gold), $\boldsymbol{\Pi}_{Tx}$ (red). Top: PIC with $\lambda = 5d_i$, $B_g = 0$. Middle: PIC with $\lambda=5d_i$, $B_g/B_m = 2.83$. Bottom: Hall-MHD with $\lambda=5d_i$, $B_g/B_m = 2.83$. Here, all quantities from the different codes have been normalised to the same Alfv\'enic units, e.g., velocities in terms of $v_{A0}$, magnetic fields in terms of $B_0$, and lengths in terms of $d_i$.}
\end{figure}

The magnetic tension term $\boldsymbol{B}\boldsymbol{B}$ has non-zero contribution only at the straight end piece of the contour, so to compare the tension more fairly between fully kinetic PIC and Hall-MHD runs we choose the length of the end piece to be the same in all calculations, $y_{\textrm{max}}= 1.6d_i$. This value is chosen so that the end contour intersects the outflow jet in both Hall-MHD and PIC. For the Harris-sheet set-up considered in Ref.~[\onlinecite{le14}] there is a significant quasi-steady period in which time derivative terms in the outflow jet region can be assumed small but, as mentioned above, for island coalescence this assumption is valid only at the peak rate. Figure~\ref{fig:wedge-contours} shows the relative contributions from the different terms in $\tensor{\boldsymbol{T}}$ to the contour integral in Eq.~(\ref{contourint}) from fully kinetic runs with $\lambda = 5d_i$, $B_g = 0$ (top) and $B_g/B_m = 2.83$ (middle), and the Hall-MHD simulation with $\lambda = 5d_i$ and $B_g/B_m = 2.83$ (bottom) at the time of the peak reconnection rate.

For the fully kinetic PIC simulation with $B_g = 0$ the inertial term is weak (blue) and the combined magnetic tension (green) and total pressure (gold) is almost entirely balanced by the stress tensor term (red). This results in an outflow velocity being much less than the upstream Alfv\'en speed in this case. Within the stress tensor contribution the majority ($85\%$) of this value is from the non-gyrotropic terms, of which the ion contribution is largest. Thus, the mechanism for which the outflow jet is slowed appears distinct from the pressure anisotropy mechanism discussed in Refs.~[\onlinecite{le14}] and~[\onlinecite{liu11}]. 

There is a significant increase in outflow inertia for the PIC simulation with $B_g/B_m = 2.83$. However, this can not be explained fully by the increase in magnetic tension ($12\%$ increase) and the reduction in the stress tensor term ($28\%$ decrease, main contribution from ion non-gyrotropic part), as the main contribution is due to a factor of three increase in $p_T$ between $B_g = 0$ and $B_g/B_m = 2.83$. This total presssure term cancels out a significant proportion of the stress tensor term, so that the magnitude of the inertial term is comparable to that of the magnetic tension. This is consistent with an outflow velocity close to the upstream Alfv\'en speed.

In contrast to the kinetic result, the total pressure and stress tensor terms are much smaller in the Hall-MHD run with $B_g/B_m = 2.83$. The Hall-MHD runs have isotropic pressure ($p_{\parallel i,e} = p_{\perp i,e}$) such that the stress tensor term is due to the collisional ion viscosity $\tensor{\boldsymbol{\Pi}}_T = \tensor{\boldsymbol{\Pi}} = -\mu \boldsymbol{\nabla} \boldsymbol{v}$. The viscous stress tensor term again mostly cancels with the total pressure so that the inertial term is balanced by magnetic tension and the outflow is Alfv\'enic.

It is clear that there are differences in the outflow physics between the fully kinetic PIC and Hall-MHD result for $B_g/B_m = 2.83$. Such differences may be related to the thickness of the ion pressure tensor sub-layer in the non-ideal electric field, shown in the bottom two panels of Fig.~\ref{fig:idr-thickness-guide}. However, these differences do not have a strong influence on the magnitude of the outflow velocity or the reconnection rate between the two runs.

\section{\label{largeguide}Effect of guide field in large systems}

\subsection{Global motions and average rates}

The results presented thus far have focussed on the smallest system-size $\lambda = 5d_i$, for which the peak rate $E_R$ differs by a factor of two between Hall-MHD and fully kinetic PIC for $B_g = 0$. However, as was demonstrated in Ref.~[\onlinecite{stanier15prl}], such order unity differences in $E_R$ can lead to different global evolution of the system. For zero guide field with $\lambda \ge 10d_i$, the separation of the island O-points in hybrid and fully kinetic PIC simulations does not monotonically decrease with time - the islands exhibit ``sloshing'' behaviour.\cite{biskamp80,craig93,knollchacon06a} When this occurs, reconnection can temporarily shut off in the periods of time that the islands move apart, resulting in a slower average reconnection rate and a longer timescale for the islands to fully merge.\cite{stanier15prl} 

\begin{figure}
\includegraphics[scale=0.9]{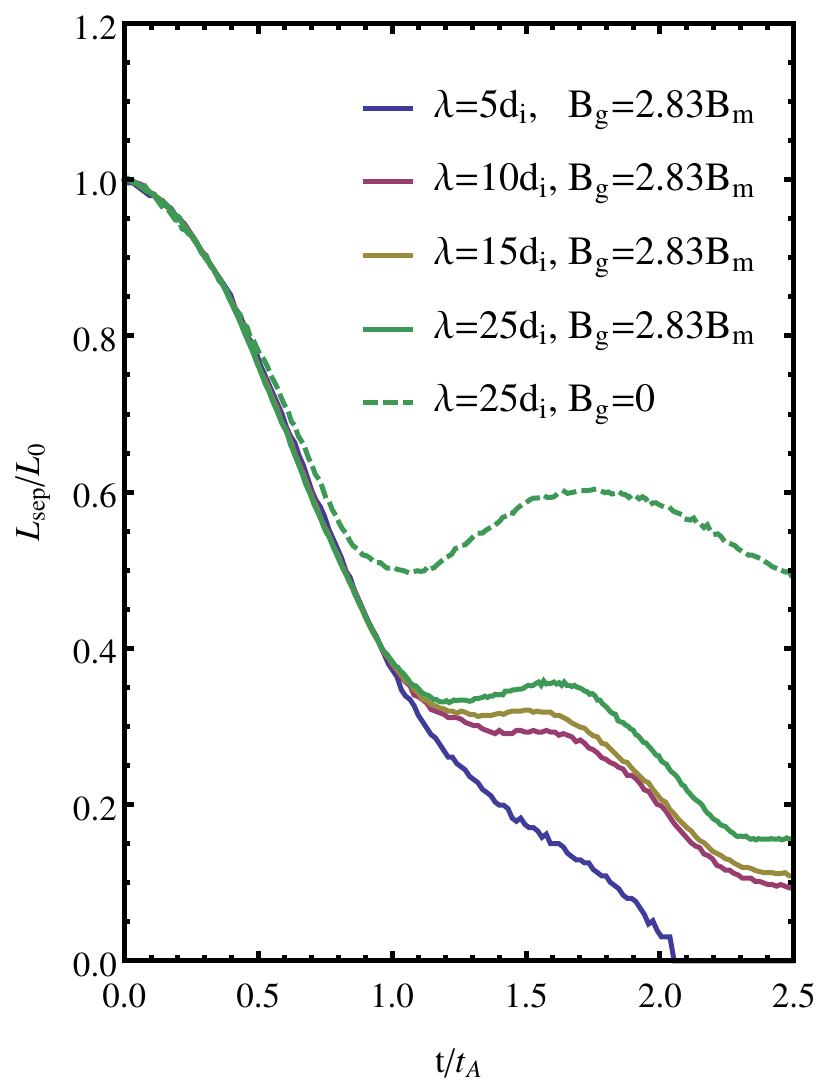}
\caption{\label{fig:opoint-gf}Normalised O-point separation distance against normalised time from PIC simulations with $B_g = 2.83B_m$ and $\lambda = (5-25) d_i$ (solid), and for $B_g = 0$ for $\lambda = 25d_i$ (dashed).}
\end{figure}

Figure~\ref{fig:opoint-gf} shows the separation between the island O-points $L_{\textrm{sep}}$, normalised by the initial separation $L_0$, versus global time for $\lambda = (5-25)d_i$ and $B_g/B_m=2.83$. The zero guide field result for the largest system size ($\lambda = 25d_i$) is also plotted for reference.~\cite{stanier15prl} Although there is reversal in the O-point separation for $\lambda \ge 10d_i$, this is significantly less than in the zero-guide field case as shown for the $\lambda=25d_i$ runs.

\begin{figure}
\includegraphics[scale=0.9]{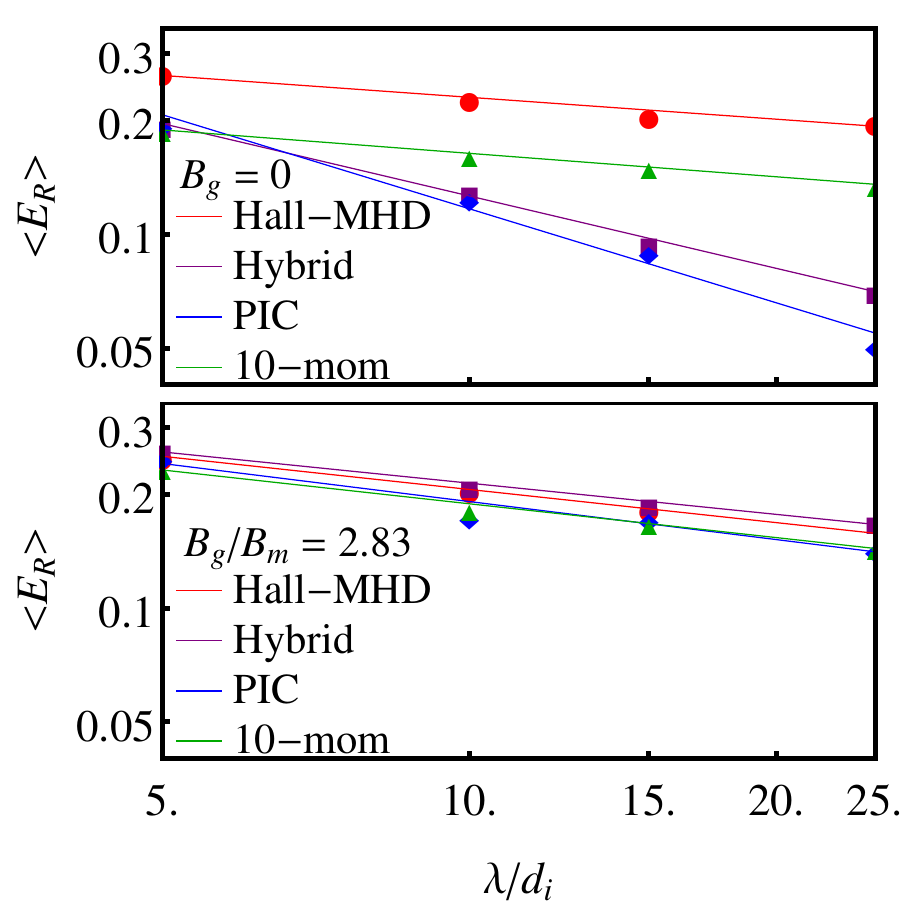}
\caption{\label{fig:average-rates} Time averaged (over $1.5\tau_A$) reconnection rate for $B_g = 0$ (top panel) and $B_g = 2.83 B_m$ (bottom panel). Results are shown from Hall-MHD (red circles), hybrid (purple squares), fully kinetic PIC (blue diamonds), and 10-moment (green triangles) simulations. The lines show the linear fits through the data. }
\end{figure}

Figure~\ref{fig:average-rates} shows the dependence of the average reconnection rates $<E_R>$ on system size, where $<>$ denotes the time average over $1.5 \tau_A$. For $B_g = 0$ (top panel) the hybrid and fully kinetic PIC simulations that include the full kinetic ion physics have an average rate that decreases steeply with system size,   $<E_R> \propto (\lambda/d_i)^{-0.8}$ for the PIC simulation, and $<E_R> \propto (\lambda/d_i)^{-0.65}$ for the hybrid simulation.~\cite{stanier15prl} This is in sharp contrast to the Hall-MHD run with $B_g=0$ which has an average rate of $<E_R> \propto (\lambda/d_i)^{-0.25}$. The 10-moment simulations agree with the fully kinetic runs for the smallest system-size ($\lambda=5d_i$) but overestimate $<E_R>$ in larger systems, where $<E_R> \propto (\lambda/d_i)^{-0.2}$, see below and Ref.~[\onlinecite{ng15}]. 
For the simulations with $B_g = 2.83 B_m$ (bottom panel), there is little difference in the average rates between the hybrid simulations with $<E_R> \propto (\lambda/d_i)^{-0.27}$, the fully kinetic PIC simulations with $<E_R> \propto (\lambda/d_i)^{-0.33}$, the 10-moment simulations with $<E_R> \propto (\lambda/d_i)^{-0.30}$, and the Hall-MHD simulations with $<E_R> \propto (\lambda/d_i)^{-0.29}$.

\begin{figure}
\includegraphics[scale=0.35]{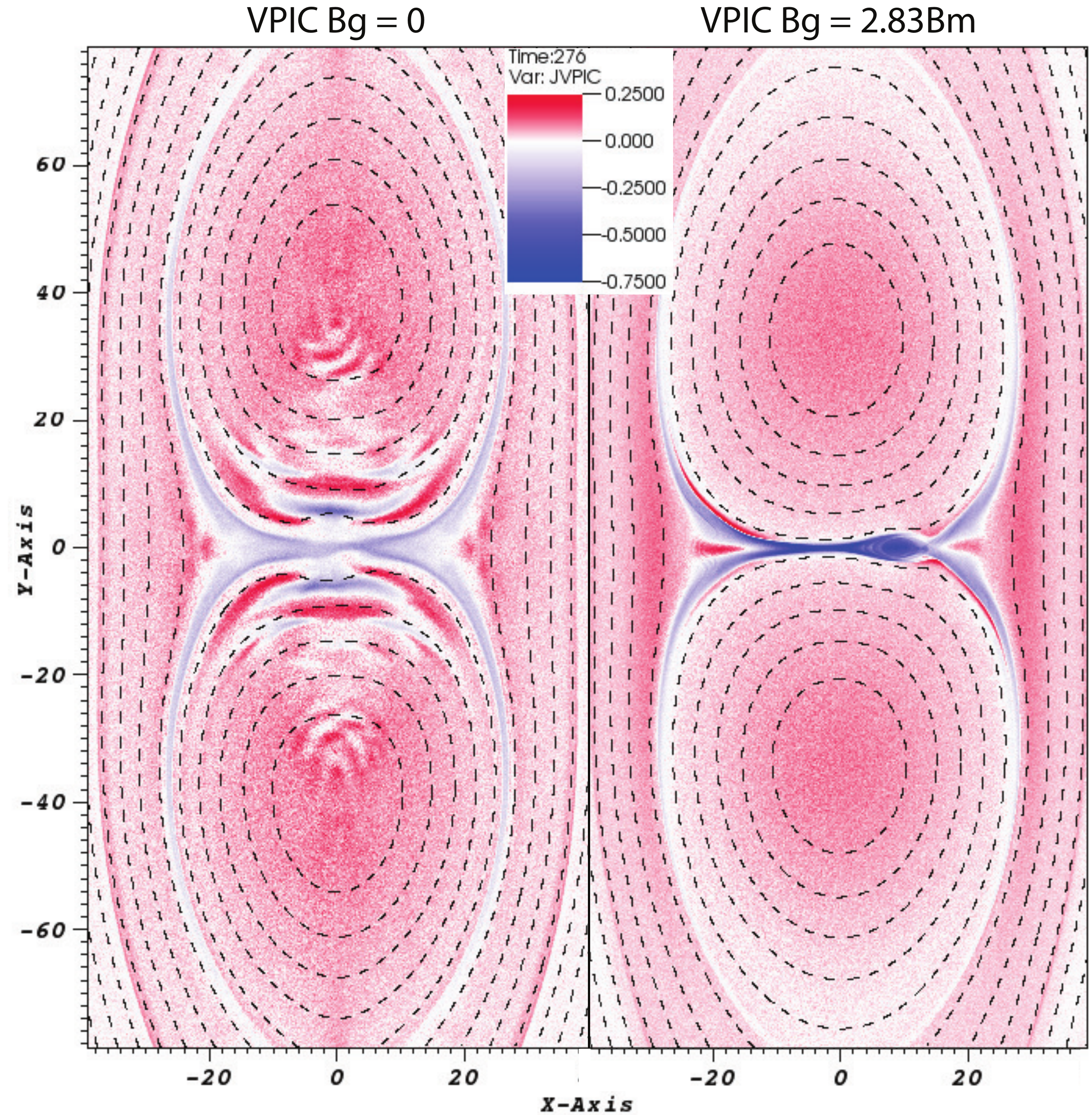}
\caption{\label{fig:island-snapshot} Current density $j_z$ in units of $B_0/\mu_0 d_i$, and contours of the magnetic potential $A_z$, from fully kinetic PIC simulations with $\lambda=25d_i$ and guide field $B_g = 0$ (left panel) and $B_g = 2.83 B_m$ (right panel). Both panels are cropped, but a movie showing the evolution within the full domain can be found in the supplementary material (Multimedia view). }
\end{figure}

\subsection{Secondary island formation}

Figure~\ref{fig:island-snapshot} (Multimedia view) shows the current density and magnetic flux from the fully kinetic PIC simulations with $\lambda=25d_i$ for the case of $B_g = 0$ (left panel) and $B_g=2.83B_m$ (right panel). When the islands collide there are significant qualitative differences in the structure of the current layer, the upstream plasma within the islands, and the outflow jets. In contrast to the zero guide-field run, the simulation with $B_g/B_m = 2.83$ has a thinner current sheet with larger current density that is unstable to the formation of secondary magnetic islands. The repeated formation and ejection of such islands is thought to be important in regulating both the length of the layer and the rate of collisionless reconnection,~\cite{daughton06} and secondary islands/plasmoids have been well studied in the limit of resistive MHD.~\cite{tajima97,loureiro07, huang13,comisso16} A complete analysis of the dynamics of these secondary islands is beyond the scope of the present study, but we report several interesting observations from these simulations. Firstly, in the case of zero guide-field we see no secondary magnetic island formation in either the Hall-MHD, hybrid or kinetic simulations for $\lambda = 5-25 d_i$. However, this does not exclude secondary islands in larger systems, and we have indeed observed a solitary island at late time ($t\approx 1.8 \tau_A$) in hybrid and fully kinetic PIC runs with $\lambda = 50 d_i$. Secondly, as discussed in Ref.~[\onlinecite{ng15}], the current layer in the zero-guide field 10-moment simulations is unstable to the formation of secondary islands for simulations with $\lambda \ge 10 d_i$. These secondary islands do appear to influence the reconnection rate, and only the smallest simulation ($\lambda = 5d_i$) without islands is found to reproduce kinetic reconnection rates - see Fig.~\ref{fig:average-rates} and Ref.~[\onlinecite{ng15}]. 

For the simulations with $B_g = 2.83 B_m$, the formation of secondary islands differs between the models. For fully kinetic PIC simulations such islands are present for $\lambda \ge 10 d_i$, but for Hall-MHD no islands are observed in the range $5-25 d_i$ and the current layer has an open X-point configuration. We note that Ref.~[\onlinecite{hirota15}] has demonstrated such an open X-point configuration is more favourable to the efficient release of magnetic energy in a reduced two-fluid model with electron diamagnetic and inertial effects. For the hybrid simulations, we also observe the open X-point configuration, and the absence of secondary magnetic islands in runs with $\lambda=5-15 d_i$. For $\lambda = 25d_i$ there are a small number of islands formed in periods where the x-point is not fully opened up. Finally, for the 10-moment simulations, secondary islands are formed in the simulations with $\lambda \geq 15d_i$. The differences in secondary island formation between these models do not strongly influence the value of $<E_R>$ for the guide field simulations, as similar values are found for all models (the bottom panel of Fig.~\ref{fig:average-rates}).

\subsection{Instability within pile-up region}

\begin{figure}
\includegraphics[scale=0.9]{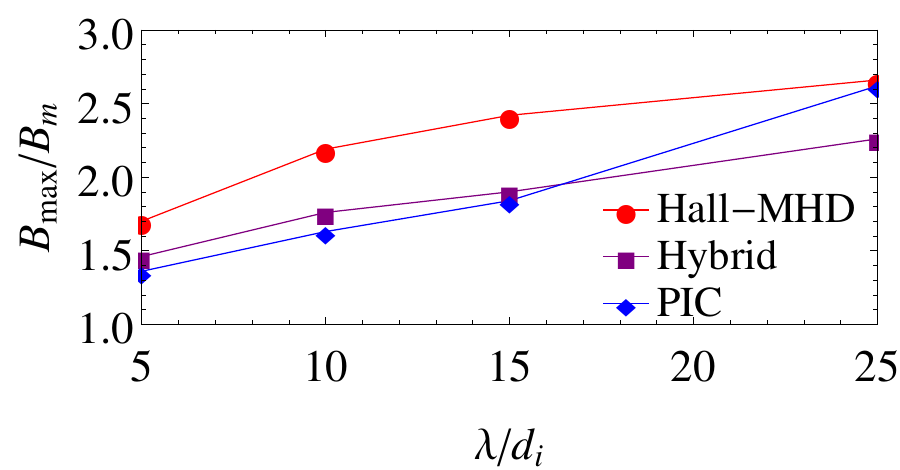}
\caption{\label{fig:pileup-gf0}Pile-up magnetic field $B_{\textrm{max}}/B_m$ against system size $\lambda/d_i$ for the zero guide-field runs. $B_{\textrm{max}}$ is the maximum value of the magnetic field strength upstream of the X-point.}
\end{figure}

It was reported in Ref.~[\onlinecite{stanier15prl}] that both hybrid and fully kinetic PIC simulations with zero guide-field have finite pressure anisotropy ($p_\perp/p_\parallel \neq 1$) within the upstream flux pile-up region at the time that the islands first collide. For the range of system sizes considered, both the maximum pile-up magnetic field strength, see Fig.~\ref{fig:pileup-gf0}, and the maximum pressure anisotropy increase with system size. 

\begin{figure}
\includegraphics[scale=0.58]{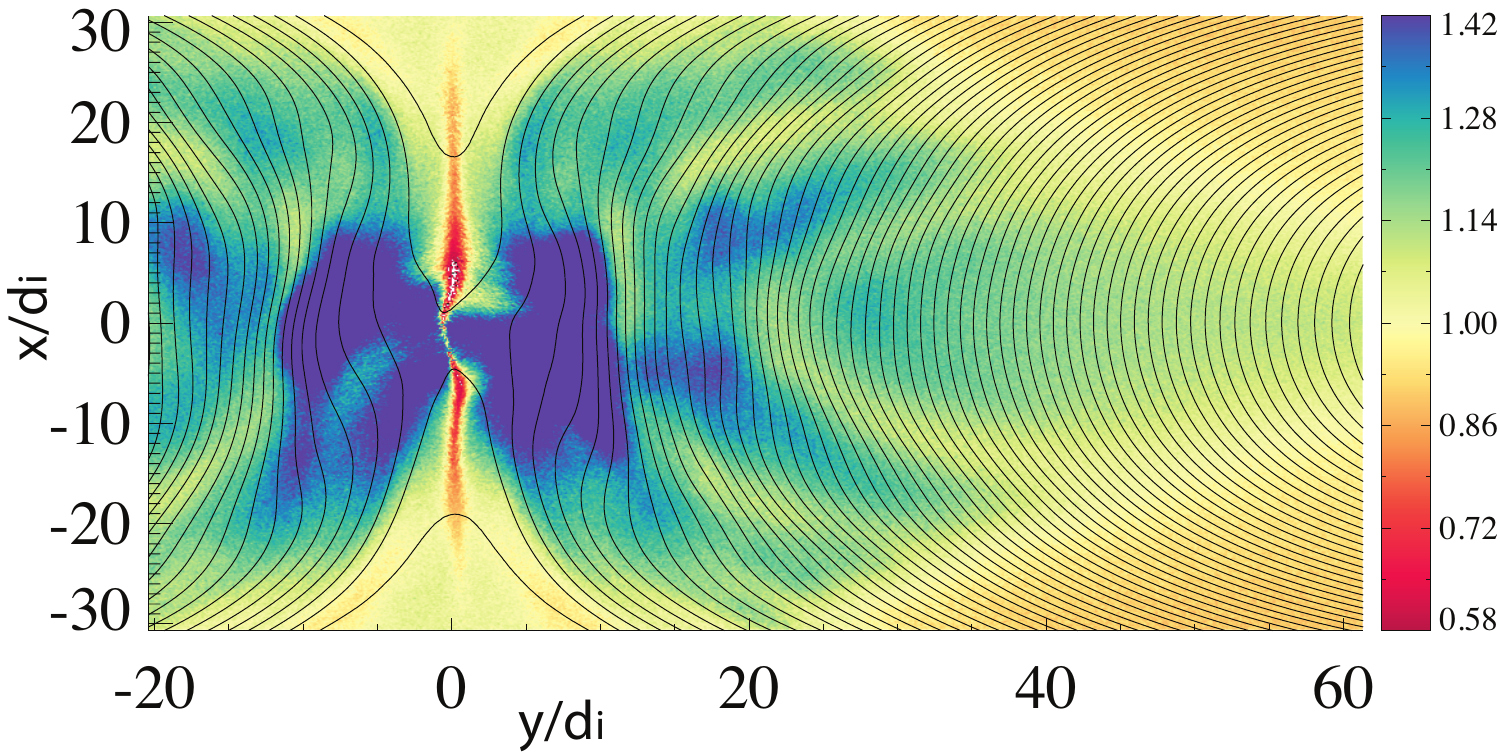}
\caption{\label{fig:ion-press-aniso}Ion pressure anisotropy $p_{i\perp}/p_{i\parallel}$ (colour scale) and magnetic field lines (lines) from VPIC simulation with $\lambda=50d_i$ and $B_g = 0$.}
\end{figure}

Figure~\ref{fig:ion-press-aniso} shows the ion pressure anisotropy $p_{\perp i}/p_{\parallel i}$ from a large fully kinetic PIC simulation with $\lambda = 50d_i$ and $B_g = 0$. There is a visible long wavelength modulation of $p_{\perp i}/p_{\parallel i}$ within a wedge shaped region of increased pressure anisotropy. A similar modulation is also visible in the current density from the PIC simulation with $\lambda=25d_i$ and $B_g = 0$ shown in Fig.~\ref{fig:island-snapshot} (Multimedia view).

Although the instability occurs in a region with finite gradients in both plasma density and magnetic field strength, we note that these gradients are larger for smaller systems in which no instability is observed. For instabilities driven by the free energy associated with the anisotropic pressure in this region ($p_{\perp}/p_{\parallel} >1$), there are three main candidate modes. In the linear regime, the mirror instability is obliquely propagating, has zero real frequency, and both ion and electron anisotropies contribute to the drive.~\cite{gary93book} The other candidates are the ion and electron cyclotron anisotropy instabilities (ICAI and ECAI), driven by the ion and electron pressure anisotropies respectively. These propagate parallel to the field in the linear regime, and have finite real frequencies. 

Fig.~\ref{fig:ion-press-aniso} shows the mode at around the time of non-linear saturation. The mode structure at this time is generally oblique to the local magnetic field direction. We also observe that the mode appears stationary with respect to the frozen-in field of the moving islands.

To further isolate the mode, we solve numerically the electromagnetic Vlasov linear dispersion relation based upon plasma parameters in the wedge shaped region at the time before the instability is visible. For simplicity, we assume a straight and uniform magnetic field, and a uniform bi-Maxwellian plasma velocity distribution. The assumption of a uniform plasma and magnetic field is not strictly valid for the region of interest, and so the calculated values should only be considered approximate.

Typical spatially averaged values of the plasma parameters within the wedge shaped region during the time leading up to the instability are $p_{i\perp}/p_{i\parallel} = 1.4$, $p_{e\perp}/p_{e\parallel} = 1.1$, $\beta_{i\parallel} = 15$, $\beta_{e\parallel} = 15$, and $v_A/c = 0.0165$ with $\beta_{i,e \parallel} = 2\mu_0 p_{i,e \parallel}/B^2$.  For these parameters, the fastest growing mode of the three is the mirror mode with growth rate $\gamma_{\textrm{mirror}}/\Omega_{ci} = 0.06$, wavenumber $kd_{i,\textrm{eff}} = 0.32$ and oblique angle to the magnetic field $\theta = 60^{\circ}$, where $d_{i,\textrm{eff}}=d_i\sqrt{n_0/n}$ is the effective ion skin depth based upon the local density. The predicted wavelength of the linear mode $\lambda_{\textrm{mirror}} = 19.6d_{i,\textrm{eff}}$ is only slightly larger than the typical measured wavelength ($\lambda \approx 12 d_i = 14.2d_{i,\textrm{eff}}$ based upon the local density $n\approx 1.4n_0$). It should be noted that these measured values are taken at non-linear saturation of the instability when the modulation is clearly visible above the background PIC noise, and the fastest growing mode from linear Vlasov theory is not necessarily the largest amplitude mode at saturation. For the stated plasma parameters, we find the ECAI is stable and the ICAI has smaller growthrate $\gamma_{\textrm{ICAI}}/\Omega_{ci}=0.03$ and longer wavelength $kd_{i,\textrm{eff}} = 0.16$.

The instability in Fig.~\ref{fig:ion-press-aniso} gives rise to local depressions (troughs) in $p_{i \perp}/p_{i \parallel}$. It also leads to strong peaks in the magnetic field strength, where the magnetic flux contours in Fig.~\ref{fig:ion-press-aniso} are close together, in similar locations to the $p_{\perp i}/p_{\parallel i}$ troughs as well as local density depressions. Such magnetic peaks are a common feature of mirror unstable plasmas in the magnetosphere, see, e.g., Ref.~[\onlinecite{soucek08}]. 

The peaks in the upstream field give rise to a significant increase in $B_{\textrm{max}}/B_m$ for the zero guide-field PIC simulation with $\lambda=25d_i$ in Fig.~\ref{fig:pileup-gf0}, since this is measured simply as the maximum upstream magnetic field. There is a much weaker increase in pile-up for the hybrid run with $\lambda=25d_i$, where signatures of this instability are very weak. The absence of the instability in this hybrid run may be due to the use of isotropic electron pressure, which means that only ion pressure anisotropy can contribute to the drive. For example, with the plasma parameters stated above but with isotropic electrons $p_{e\perp}/p_{e\parallel} =1$ the growth rate is reduced by almost a factor of 2 to $\gamma_{\textrm{mirror}}/\Omega_{ci} = 0.034$. For a larger hybrid run with $\lambda=50d_i$ we do see structures in $p_{i \perp}/p_{i \parallel}$ and the magnetic field that are very similar to those in Fig.~\ref{fig:ion-press-aniso}. The difference in system-size threshold at which the instability is clearly observed between hybrid and PIC also coincides with a greater difference in the reconnection rate between the two codes. In particular, Fig.~7 shows a significantly lower average reconnection rate in the $\lambda=25d_i$ and $B_g=0$ PIC run where the instability is present, compared with the hybrid result for the same system size. If the average reconnection rate for the $B_g = 0$ PIC simulations is calculated neglecting the $\lambda=25d_i$ run, the system-size scaling is $<E_R> \propto (\lambda/d_i)^{-0.69}$ which is much closer to the hybrid scaling. It is conceivable that the presence of this mirror-like instability upstream of the reconnection site leads to further reductions in the reconnection rate in large systems, however, further study is needed to address this issue.

\section{\label{conclusions}Conclusions}

Simulations of magnetic reconnection using the island coalescence problem set-up naturally incorporate several key features of real reconnecting systems that are absent in simpler elongated current layer models. In particular, the island coalescence problem includes the self-consistent formation of the current layer as the magnetic islands collide, and the two-way coupling between the micro-physics of the diffusion region and the macro-scale system that supplies the magnetic flux. For accurate modelling of near Earth space weather, where global fully kinetic simulations are currently impractical, it is important to understand the accuracy and limitations of reduced physics models for the simulation of such systems.

Refs.~[\onlinecite{karimabadi11}],~[\onlinecite{stanier15prl}], and~[\onlinecite{ng15}] found significant differences between reconnection during island coalescence and earlier current layer studies in the zero guide-field limit. The Hall-MHD model, which had been thought sufficient to reproduce the reconnection rates of fully kinetic simulations,~\cite{birn01} was found inadequate to describe the kinetic results for the island coalescence problem. The Hall-MHD model overestimates the reconnection rate, is unable to reproduce its dependence on ion temperature or system-size, and gives incorrect values for the pile-up magnetic field strength, outflow velocity, and ion diffusion region thickness.~\cite{stanier15prl} 

The hybrid model, with massless fluid electrons and fully kinetic ions, was demonstrated as the minimum sufficient model to reproduce the above features of the island coalescence problem in the zero guide field limit.~\cite{stanier15prl} The 10-moment fluid model was found in good agreement with kinetic results for the smallest system-size, but for larger systems is unstable to the formation of many secondary magnetic islands which are not present in the full kinetic simulations.~\cite{ng15}

In this study we considered the sensitivity of these results to the addition of a finite guide field, since this can modify the kinetic ion physics that is crucial in setting the reconnection rate. We consider a range of guide fields $B_g/B_m = 0 - 5.66$ that is suitable for magnetospheric applications. Unexpectedly it is found that for a limited range of guide fields ($B_g/B_m < 2.83$) the reconnection rate, outflow velocity and pile-up field strength increase with increasing $B_g$ in the full kinetic, hybrid and 10-moment models. The same quantities for Hall-MHD vary little over this range of $B_g$, and for $B_g/B_m \ge 2.83$ the values of these quantities are in good agreement between the Hall-MHD and fully kinetic simulations. To understand this behaviour, the physics breaking the ion frozen-in condition and contributing to the outflow jet momentum balance was considered.

Only for the large guide fields $B_g/B_m \geq 2.83$ is there qualitative agreement in the physics setting the ion diffusion region thickness, where ion inertia is the first term to break the frozen-in condition for ions in both the Hall-MHD and kinetic models. At the X-point, this guide field is sufficient to suppress meandering motion, so that the distribution function becomes to lowest order gyrotropic around the guide field direction. The outflow jet speed is reduced significantly with respect to the upstream ion Alfv\'en speed $v_{A,in}$ in the zero guide-field kinetic case due to a quasi-viscous effect associated with the non-gyrotropic part of the ion pressure tensor. For $B_g/B_m = 2.83$ there are still significant differences in the strength of the ion stress tensor and total pressure contributions to momentum balance between Hall-MHD and kinetic, but these terms mostly cancel so that they do not influence the outflow speed. The magnetic tension is balanced by ion inertia to give an outflow speed comparable to $v_{A,in}$. 

For the island coalescence problem, small changes of the peak reconnection rate can have larger consequences on the evolution of the system. If the reconnection rate is too slow, the islands can bounce off each other and reconnection can be temporarily switched off until the islands come together again. It is found that in large kinetic systems, islands with strong guide field have larger average reconnection rates and merge on a faster timescale than those with zero guide field. In addition to the changes in outflow velocity and ion diffusion region geometry with guide field that are present in small systems, large fully kinetic systems with guide field are also unstable to repeated formation of secondary magnetic islands which may play a role in limiting the length of the reconnection layer. Also, the presence of a mirror-like instability found in the largest zero guide field kinetic systems may further reduce the reconnection rate, but further work is needed to confirm this conclusion. This instability is not present in the strong guide field runs.

Reconnection events at the Earth magnetosphere can have a wide range of guide fields, but in the Earth's magnetotail it is often small~\cite{eastwood10prl} compared with the reconnecting field. Indeed, many of the early observations from the Magnetospheric Multiscale Mission have found reconnection events with weak guide field at the magnetopause, see e.g. Ref.~[\onlinecite{burch16}]. This study, along with Refs.~[\onlinecite{stanier15prl}] and~[\onlinecite{ng15}], demonstrate the limitations of the Hall-MHD model for reconnection events in such locations. The hybrid model, with massless fluid electrons and kinetic ions, and the 10-moment fluid model both give results in better agreement with the full kinetic system.

\begin{acknowledgements}This work is supported by the U.S. Department of Energy, Office of Science, and Office of Fusion Energy Sciences, and by the Collaborative Space Weather Modeling Program through NSF Grant No. AGS-1338944 and NASA Grant No. NNH13AW51I. The work used resources provided by the Los Alamos National Laboratory Institutional Computing Program, which is supported by the U.S. Department of Energy National Nuclear Security Administration under Contract No. DE-AC52-06NA25396.\end{acknowledgements}


%

\end{document}